\documentclass[superscriptaddress,aps,nofootinbib,twocolumn]{revtex4}
\usepackage{amsfonts}
\usepackage{amsmath}
\usepackage{amssymb}
\usepackage{graphicx}

\begin{document}
\preprint{UCL-IPT-03-09}
\title{Polychromatic Grand Unification}
\author{G. C. Branco}
\affiliation{Departamento de F\'{\i}sica and Grupo Te{\'o}rico de F{\'\i}sica de
Part{\'\i}culas, Instituto Superior T\'{e}cnico, Av. Rovisco Pais, 1049-001 Lisboa,
Portugal}
\author{J.-M. G{\'e}rard}
\affiliation{Institut de Physique Th{\'e}orique, Universit{\'e} Catholique de Louvain,
Chemin du Cyclotron 2, B-1348 Louvain-la-Neuve, Belgium}
\author{R. Gonz\'alez Felipe}
\affiliation{Departamento de F\'{\i}sica and Grupo Te{\'o}rico de F{\'\i}sica de
Part{\'\i}culas, Instituto Superior T\'{e}cnico, Av. Rovisco Pais, 1049-001 Lisboa,
Portugal}
\author{B. M. Nobre}
\affiliation{Departamento de F\'{\i}sica and Grupo Te{\'o}rico de F{\'\i}sica de
Part{\'\i}culas, Instituto Superior T\'{e}cnico, Av. Rovisco Pais, 1049-001 Lisboa,
Portugal}

\begin{abstract}
The possibility of achieving a gauge coupling grand unification by increasing
the QCD degrees of freedom at higher energies is investigated. When confronted
with precision low-energy data, the generic weak-mixing angle relation
$\sin^2\theta_W = N/2(N+1)$ requires $N=7$ colors at the GUT scale $M_U \simeq
10^{17}$~GeV. With the exclusive addition of light Higgs doublets to the
standard model particle content, such a unification may also occur if $N=5$ at
$M_U \simeq 10^{16}$~GeV.
\end{abstract}

\date{\today}
\maketitle

\section{Introduction}

The possible merger of the standard model (SM) gauge couplings at some high
energy scale is a very attractive scenario in our quest for unification
\cite{Witten:2002ei}. It is well known that precision low-energy data rule out
minimal grand unified theories (GUT) with the SM particle content. On the other
hand, unification of gauge couplings can be achieved within the framework of
\emph{supersymmetric extensions of the SM}  \cite{Amaldi:1991cn}. Therefore, it
is tempting to consider the above as a strong indication for low-energy
supersymmetry. Nevertheless, given the theoretical and experimental
implications of this unification scheme, it is also pertinent to ask whether
other scenarios might lead to unification of gauge couplings.

As a matter of fact, the apparent absence of light exotic fermions is nowadays
an argument in favor of the weak-mixing relation $\sin^2\theta_W = 3/8$, linked
to the well-known chain of simple group embeddings $SU(5) \subset SO(10)
\subset E_6 \subset \ldots$. In this letter, we shall consider other group
embeddings within the framework of \emph{polychromatic extensions of the SM}.

The possibility that the relevant number of colors $N$ increases with energy
has indeed been considered by 't Hooft as a requirement for obtaining not only
asymptotic freedom but also ``asymptotic convergence" at high energies
\cite{'tHooft:1982di}. Moreover, it is also quite remarkable that a simple
polychromatic extension of the SM, i.e. $SU(N)_c \times SU(2)_L \times
U(1)_{1/N}$, is still free of ``triangular" \cite{Bardeen:md} and ``mixed"
\cite{Alvarez-Gaume:1983ig} gauge anomalies within one family of left-handed
quarks and leptons displayed as follows:
\begin{align} \label{SUNfund}
\left(
\begin{array}{c}
  u \\
  d \\
\end{array}
\right)_{1/N} \quad \left(
\begin{array}{c}
  \nu_e \\
  e \\
\end{array}
\right)_{-1} \quad (u^c)_{-1-1/N} \quad (d^c)_{1-1/N} \quad (e^c)_2\quad,
\end{align}
the subscript being the hypercharge quantum number $Y$.  As in the SM, each
family is distributed over the five multiplets $(\pmb{N},\pmb{2})$, $2 \times
(\pmb{N},\pmb{1})$, $(\pmb{1},\pmb{2})$ and $(\pmb{1},\pmb{1})$ of $SU(N)_c
\times SU(2)_L$. The value of $\sin^2\theta_W$ at the unification scale may be
computed by requiring the one-loop fermionic contribution to the $\gamma - Z$
mixing diagrams to be finite \cite{Branco:1982qi}, which implies the relation
\begin{equation} \label{tracecond}
  \mbox{Tr}\left[\,Q\left(T_{3L}-Q \sin^2\theta_W\right)\right]=0\,,
\end{equation}
where the trace is computed over \emph{all} the fermions in the theory and the
charge operator is given by the relation $Q=T_{3L} + Y/2 $. In the case of the
charge assignment (\ref{SUNfund}), it leads to
\begin{align} \label{sin2theta}
  \sin^2\theta_W=\frac{N}{2(N+1)}\,.
\end{align}

For the SM fermionic content ($N=3$), we recover the standard value
$\sin^2\theta_W = 3/8$ which naturally follows from the $SU(5)$ or $SO(10)$
unification groups. Remarkably, we shall see that Eq.~(\ref{sin2theta}) turns
out to be also a generic relation for the weak-mixing angle in polychromatic
$SU(N+2)$ or $SO(2N+4)$ unification groups. The corresponding value of
$\sin^2\theta_W$ is surprisingly stable with respect to variations of $N$ from
three to infinity. However, low-energy data on gauge couplings already provide
strong restrictions on the value of $N$ at the GUT scale. In fact, gauge
coupling grand unification is only possible in two cases, namely, for $(N=7,
n_H=1)$ and $(N=5, n_H=3)$, if $n_H$ is the number of light Higgs doublets.

\section{Beyond $\pmb{\sin^2\theta_{W}=3/8}$}

The angle $\theta_W$ defines both the tree-level gauge-boson eigenstates
($m_W^2/m_Z^2=\cos^2 \theta_W$) and the ratio of weak gauge couplings ($\tan
\theta_W= g^\prime/g_2$).

On one hand, if the complete fermion content of the GUT is known, the value of
$\sin^2\theta_W$ at the gauge coupling unification scale may still be computed
by imposing the cancellation of divergencies in the one-loop fermionic
contribution to the $\gamma-Z$ mixing. For illustration, let us consider the
Pati-Salam model \cite{Pati:1974yy} recently revived in the context of brane
models \cite{Leontaris:2000hh}. For an arbitrary number of colors $N$, we need
the following two multiplets:
\begin{equation}\label{PS}
  F^e_{L,R}=\left[\begin{array}{cccc}
    u_1 & ... & u_N & \nu_e \\
    d_1 & ... & d_N & e \\
  \end{array}\right]_{L,R},
\end{equation}
which transform as $(\pmb{N+1,2,1})$ and $(\pmb{\overline{N+1},1,2})$ under
$SU(N+1)_c \times SU(2)_L \times SU(2)_R$. The electric charge operator is now
given by $ Q = T_{3L} + T_{3R} + (B-L)/2 $, where $B, L$ are the baryon and
lepton numbers, respectively. This relation leads then to the charges
\begin{align}\label{PScharges}
Q_u = \frac{1+N}{2N}\,, \quad Q_d = \frac{1-N}{2N}\,,
\end{align}
already predicted by the anomaly-free $SU(N)_c \times SU(2)_L \times
U(1)_{1/N}$ model (cf. Eq.~(\ref{SUNfund})), and subsequently, to the same
value of $\sin^2\theta_W$ (cf. Eq.~(\ref{sin2theta})) since the additional
right-handed neutrinos do not couple to the photon.

On the other hand, if we only know the gauge group, the corresponding value of
the weak-mixing angle at the GUT scale can be directly derived from the
properly normalized weak gauge couplings. As is well known, the minimal simple
group in which $SU(3)_c \times SU(2)_L \times U(1)_Y$ can be embedded is
$SU(5)$, with the SM fermions accommodated in the $\pmb{\bar{5}}$ and $(\pmb{5}
\otimes \pmb{5})_a = \pmb{10}$ representations \cite{Georgi:sy}. We may
generalize this procedure for $N > 3$ in order to embed $SU(N)_c \times SU(2)_L
\times U(1)_{1/N}$ into a simple group. The minimal embedding is obviously
$SU(N+2)$ with the traceless hypercharge generator associated with $U(1)_{1/N}$
given by
\begin{align}
\left(\frac{N}{N+2}\right)^{1/2} \text{diag}\, \left( \frac{2}{N}\, , \,
\frac{2}{N}\, ,\, \ldots\,,\, \frac{2}{N}\, , \, -1\, , \, -1 \right)\,,
\end{align}
and properly normalized so that $\text{Tr}\,\left(Y^2\right)=2$. In this case,
new (left-handed) colored particles with charge $1/N$ are needed once $N>3$.
However, from the normalization condition $g^2_1= g'^2 (N+2)/N\,$ and the
boundary condition $g_3=g_2=g_1$ on the $SU(N)_c \times SU(2)_L \times
U(1)_{1/N}$ gauge couplings, we get again the GUT relation
\begin{align}
 \sin^2\theta_W \equiv \frac{g'^2}{g^2_2+g'^2}=\frac{N}{2(N+1)}\,.
\end{align}

We remark that in order to preserve the gauge anomaly cancellation, we must
include more than one irreducible representation of the group (which can always
be obtained by tensor products of the fundamental representation). All the
particle charges in each representation are uniquely fixed and may be obtained
from the charges of the fundamental representation. It is important to note
that each irreducible representation leads to the same value of
$\sin^2\theta_W$, provided that all the particles in the representation are
included in the trace of Eq.~(\ref{tracecond}). This means that the fundamental
representation of $SU(N+2)$ alone fixes the value of $\sin^2\theta_W$.
Furthermore, this result is independent of the way the SM-like particles are
distributed among the representations. For illustration, let us consider the
following two simple anomaly free realizations.

It is known that an $SU(N+2)$ gauge theory built from $(N-2)\,\times
\overline{(\pmb{N+2})}$ and $\left[(\pmb{N+2})\otimes(\pmb{N+2})\right]_a $
matter field representations is anomaly free. The decomposition of the
fundamental representation $(\pmb{N+2})$ under $SU(N)_c \times SU(2)_L \times
U(1)_{1/N}$ is $(\pmb{N},\pmb{1},2/N) \oplus (\pmb{1},\pmb{2},-1)$, which
allows us to identify the $N$-colored quarks and their chiral charges:
\begin{align}\label{chiralcharges}
  Q_{u_L} = 1-\frac{1}{N}\,, \quad Q_{u_R} = \frac{2}{N}\,, \quad
  Q_{d_L} = Q_{d_R} = -\frac{1}{N}\,.
\end{align}
The remaining colored particles are quite exotic.

If we assume instead that no irreducible representation appears more than once
within one family, then, in order to preserve the gauge anomaly cancellation,
we need the following set of fermion irreducible representations
\cite{Georgi:1979md}
\begin{align}\label{Georgi}
  \sum_{l=1}^{(N+1)/2} [\underbrace{(\pmb{N+2}) \otimes \cdots
  \otimes (\pmb{N+2})}_{2l}]_a\, \in\, SU(N+2)\, ,
\end{align}
obtained by antisymmetrizing tensor products of the fundamental representation.
In such a case, the charges of the $N$-colored quarks are vector-like:
\begin{align}\label{vectorcharges}
  Q_{u_L}=Q_{u_R}=1-\frac{1}{N}\,, \quad \quad Q_{d_L}=Q_{d_R}=-\frac{1}{N}\,,
\end{align}
and the fermion color is restricted to $\pmb{1}$, $\pmb{3}$ and $\bar{\pmb{3}}$
of $SU(3)_c$ \cite{Gell-Mann:1976pg}.

From Eq.~(\ref{Georgi}) it is then tempting \cite{Fritzsch:nn} (natural, if
right-handed neutrinos are included) to consider the $\pmb{2^{N+1}}$
dimensional spinor representation of the larger group $SO(2N+4)$.

In summary, we have at our disposal a generic value of $\sin^2\theta_W$ which
can be motivated by various polychromatic unification scenarios based on the
gauge group $G(N)= SU(N+1) \times SU(2)_L \times SU(2)_R\,$, $SU(N+2)$ or
$SO(2N+4)$, characterized by different charge assignments (see
Eqs.~(\ref{PScharges}), (\ref{chiralcharges}) and (\ref{vectorcharges}),
respectively) if $N>3$. However, these gauge groups are just examples among
others. For our purposes, all the crucial information is encoded in relation
(\ref{sin2theta}), which we shall now assume to be valid at the GUT scale.

Finally, we emphasize that the number $N$ of ``active" colors at the GUT scale
must be odd to ensure

\noindent - a cancellation of the ``global" $SU(2)$ anomaly \cite{Witten:fp}
for the $SU(N+1) \times SU(2)_L \times SU(2)_R\,$ extended Pati-Salam
unification (see Eq.~(\ref{PS}));

\noindent - the natural embedding of $SU(N+2)$ into $SO(2N+4)$ (see
Eq.~(\ref{Georgi})).

\section{Gauge coupling unification}

If the GUT scale $M_U$ is large compared with all the SM particle masses, the
one-loop renormalization group equations (RGE) for the three gauge coupling
constants can be written in the form \cite{Georgi:yf}
\begin{align} \label{RGE}
   \alpha^{-1}_i(M_U)=\alpha^{-1}_{iZ} - \frac{b_i}{2\pi}\ln{\frac{M_U}{m_Z}}\,,
\end{align}
where $\alpha_i=g_i^2/(4\pi)$ and  $\alpha_{iZ} \equiv \alpha_i(m_Z)$. The
coefficients $b_i$ are gauge group dependent. From Eqs.~(\ref{RGE}), it is
straightforward to show that for the gauge couplings $g_i$ to merge at $M_U$,
the values of $\alpha^{-1}_{iZ}$ and $b_i$ must obey the relation
\begin{align} \label{frelation}
 f_b = f_\alpha\,,
\end{align}
with
\begin{align} \label{falphafb}
f_b \equiv \frac{b_1-b_3}{b_2-b_3} \quad , \quad f_{\alpha} \equiv
\frac{\alpha^{-1}_{3Z}-\alpha^{-1}_{1Z}}{\alpha^{-1}_{3Z} -\alpha^{-1}_{2Z}}\,.
\end{align}
For instance, assuming that the standard $SU(5)$ or $SO(10)$ unification group
breaks into $SU(3)_c \times SU(2)_L \times U(1)_Y$ at the GUT scale, one has
the values
\begin{align} \label{biSM}
b_i^{(SM)} = \left(\frac{41}{10},-\frac{19}{6},-7\right)\,,
\end{align}
which imply $f_b^{(SM)} \simeq 2.9$. Using the following precision measurements
as inputs \cite{Hagiwara:pw}
\begin{align} \label{ewdata}
\alpha^{-1}(m_Z) &=127.934 \pm 0.027\,, \nonumber\\
\sin^2 \theta_W (m_Z) &= 0.23113 \pm 0.00015\,, \\
\alpha_s(m_Z) &= 0.1172 \pm 0.0020\,, \nonumber
\end{align}
which are defined in the $\overline{MS}$ scheme, one obtains $f_\alpha \simeq
2.4$. This means that the gauge coupling constants do not unify in the SM
embedded into $SU(5)$ or $SO(10)$ gauge groups. On the other hand, in the
minimal supersymmetric extension of the SM (MSSM) where
\begin{align}
b_i^{(MSSM)} =\left(\frac{33}{5},1,-3\right)\,,
\end{align}
one obtains $f_b^{(MSSM)}=2.4$, i.e an almost perfect gauge coupling
unification.

We wonder whether it is possible to obtain a gauge coupling unification in the
context of $G(N)$ gauge groups. Let us assume that the relevant $G(N)$ is
directly broken into $SU(3)_c \times SU(2)_L \times U(1)_Y$ at the GUT scale
$M_U$ and that, besides the SM particle content, only new Higgs doublets may
acquire their masses at the electroweak scale. Then, the one-loop RGE are still
given by Eq.~(\ref{RGE}), where now the coefficients $b_i$ read
\begin{align} \label{biSMnH}
b_i(N,n_H) =
\left(\frac{40+n_H}{6}\frac{N}{N+2},-\frac{10}{3}+\frac{n_H}{6},-7\right)\,,
\end{align}
with $n_H$ the total number of light Higgs doublets. Once again, to achieve
unification the relations (\ref{frelation}) and (\ref{falphafb}) must hold with
\begin{align} \label{alpha12Z}
\alpha_{1Z} = \frac{\alpha(m_Z)}{1-\sin^2\theta_W(m_Z)} \frac{N+2}{N}\,, \quad
\alpha_{2Z} = \frac{\alpha(m_Z)}{\sin^2\theta_W(m_Z)}\,.
\end{align}

\begin{figure}
\begin{center}
\includegraphics[width=8cm]{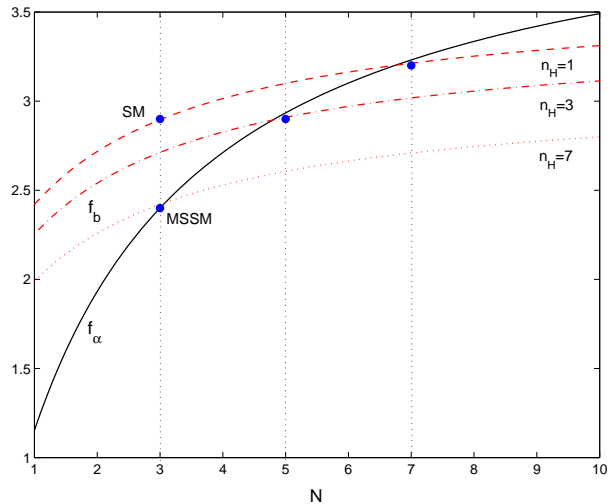}
\caption{The parameters $f_\alpha$ (solid curve) and $f_b$ (dotted and dashed
curves) as functions of the number of colors $N$ and light Higgs doublets $n_H$
. The function $f_\alpha$ is plotted using the central values of the couplings
$\alpha_i(m_Z)$ as given in Eqs.~(\ref{ewdata}).} \label{fig1}
\end{center}
\end{figure}

In Fig.~\ref{fig1} we plot the functions $f_\alpha$ and $f_b$ as functions of
the number of colors $N$ for $n_H=1,3,7$, which turn out to be the only
possible solutions compatible with gauge coupling unification. For comparison,
the SM and MSSM solutions are also indicated. In fact, the drastic
supersymmetric extension of the SM causes only small changes in the predicted
relation among gauge couplings. The reason is that any addition of particles
forming complete GUT multiplets will cause small changes if they are (nearly)
degenerate in mass. So the major difference arises from the fact that the MSSM
requires 2 Higgs doublets which do not have accompanying light triplets. Taking
into account their fermionic superpartners, we understand why it almost
coincides with the (non-supersymmetric) $N=3$ solution with approximately $6 =
2\times(1+2)$ scalar doublets.

\begin{figure}
\begin{center}
\includegraphics[width=8.cm]{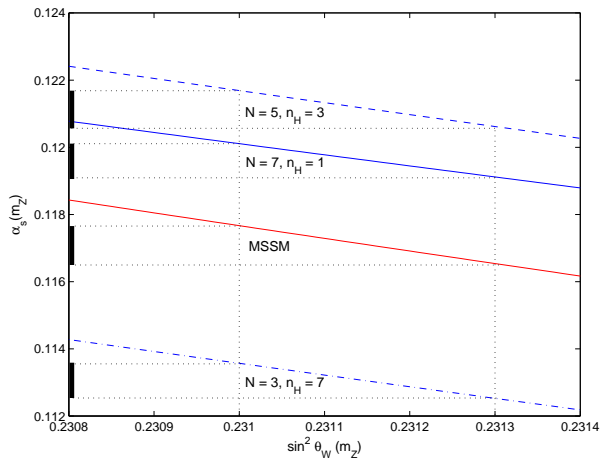}
\caption{The strong coupling constant $\alpha_s(m_Z)$ as a function of $\sin^2
\theta_W(m_Z)$ for the three possible solutions $(N=7,n_H=1),\ (N=5,n_H=3),\
(N=3,n_H=7) $, as well as for the MSSM. The vertical dotted lines corresponds
to the experimentally allowed range as given in Eqs.~(\ref{ewdata}). The black
bars on the vertical axis represent the corresponding theoretically allowed
ranges for $\alpha_s(m_Z)$ to ensure gauge coupling unification.} \label{fig2}
\end{center}
\end{figure}

To check the consistency of these solutions with the electroweak precision
measurements given in Eqs.~(\ref{ewdata}), we present in Fig.~\ref{fig2} the
strong coupling constant $\alpha_s(m_Z)$ as a function of $\sin^2
\theta_W(m_Z)$, the latter being more precisely measured. We notice that the
present uncertainties on $\alpha_s(m_Z)$ do not allow us to exclude any of the
four solutions. Yet the bounds on proton decays provide us with another way to
disentangle them.

The unification scale $M_U$ is easily obtained from Eqs.~(\ref{RGE}),
(\ref{biSMnH}) and (\ref{alpha12Z}):
\begin{align} \
M_U = m_Z \exp \left[ \frac{2 \pi(\alpha^{-1}_{1Z}-\alpha^{-1}_{2Z})}{b_1-b_2}
\right]\,.
\end{align}
In particular, for the solutions previously found, the unification occurs at
\begin{align}
N=7 &\quad n_H=1 \quad M_U \simeq 1.2 \times 10^{17}~\text{GeV}\,,\nonumber\\
N=5 &\quad n_H=3 \quad M_U \simeq 8.4 \times 10^{15}~\text{GeV}\,,\nonumber\\
N=3 &\quad n_H=7 \quad M_U \simeq 4.6 \times 10^{13}~\text{GeV}\,.
\end{align}
For the MSSM solution we obtain the well-known value $M_U \simeq 2.0 \times
10^{16}$~GeV.

We notice that the unification scale in the case of $(N=3,n_H=7)$ is obviously
too low to be compatible with the experimental constraints on proton decay
\cite{Shiozawa:1998si}. Consequently, we are left with the polychromatic cases
$N=7$ and $N=5$. As a matter of fact, the MSSM with a GUT scale around
$10^{16}$~GeV might already be ruled out by proton decay $(p\rightarrow
\bar{\nu}\, K^{+})$ data from Super-Kamiokande \cite{Murayama:2001ur}.

\section{Conclusion}

Polychromatic unification is a possible alternative to supersymmetric
unification. In the minimal $N=7$ polychromatic $SU(9)$ or $SO(18)$ GUT
\cite{Frampton:1979fd}, the unification scale around $10^{17}$~GeV is high
enough since the proton lifetime $(p\rightarrow e^{+}\, \pi^{0})$ is
generically proportional to the fourth power of the superheavy gauge boson
masses \cite{Georgi:yf}. Moreover, $10^{17}$~GeV is close enough to the Planck
scale of $10^{19}$~GeV to suggest a unification with gravity
\cite{Witten:2002ei}.

The two polychromatic solutions ($N=7$ and $N=5$) presented here are based on
the hypothesis of a direct symmetry breakdown of the relevant gauge group
$G(N)$ into the standard $SU(3)_c \times SU(2)_L \times U(1)_Y$ at a unique GUT
scale. However, stepwise symmetry breakdown from high to lower $N$ could also
take place at nondegenerate high scales, possibly giving rise to new solutions
with $N>7$, in agreement with 't Hooft ``suspicion" that at infinite energy,
$N$ has to tend to infinity \cite{'tHooft:1982di}. But these threshold effects
require two-loop RGE which are clearly beyond the scope of this letter.

\begin{acknowledgments}
This work was financed by {\em Funda\c{c}{\~a}o para a Ci{\^e}ncia e a Tecnologia} (FCT,
Portugal) through the projects POCTI/FNU/43793/2002 and CFIF - Plurianual
(2/91). The work of J.-M.G. has been partially supported by the Federal Office
for Scientific, Technical and Cultural Affairs through the Interuniversity
Attraction Pole P5/27.  The work of R.G.F. and B.M.N. was supported by FCT
under the grants SFRH/BPD/1549/2000 and SFRH/BD/995/2000, respectively.
\end{acknowledgments}

\end{document}